\documentstyle[11pt]{article}
\begin{document}
\setcounter{page}{1}
\pagestyle{plain}
\setcounter{equation}{0}
%
%
%
\ \\[12mm]
\begin{center}
    {\bf ON MATRIX PRODUCT GROUND STATES FOR\\[1mm]
       REACTION-DIFFUSION MODELS}
	     \\[20mm]
\end{center}
\begin{center}
\normalsize
	Haye Hinrichsen$^{\diamond}$,
        Sven Sandow$^{\diamond}$
        and Ingo Peschel$^{\star,}$\footnote{Permanent address:
        Freie Universit\"{a}t Berlin,
        Institut f\"ur Theoretische Physik, Arnimallee 14,\\ .\hspace{33mm}
        D-14195 Berlin, Germany}\\[13mm]
	$^{\diamond}$ {\it Department of Physics of Complex Systems\\
	Weizmann Institute of Science\\ Rehovot 76100, Israel}\\[4mm]
	$^{\star}$ {\it Max-Planck-Institut f\"ur
        Physik komplexer Systeme\\
        Bayreuther Str. 40, Haus 16\\
        01187 Dresden, Germany}
\end{center}
\vspace{2cm}
{\bf Abstract:}
We discuss a new mechanism leading to a matrix product form
for the stationary state of one-dimensional stochastic models.
The corresponding algebra is quadratic and involves four different
matrices. For the example of a coagulation-decoagulation
model explicit four-dimensional representations are given
and exact expressions for various physical quantities are  recovered.
We also find the general structure  of $n$-point correlation functions
at the phase transition.
\\[30mm]
cond-mat/9510104
\\[3mm]
\rule{7cm}{0.2mm}
\begin{flushleft}
\parbox[t]{3.5cm}{\bf Key words:}
\parbox[t]{12.5cm}{Reaction-diffusion systems, matrix product ground states,
		   coagulation model, integrability}
\\[2mm]
\parbox[t]{3.5cm}{\bf PACS numbers:} 02.50.Ey, 05.60.+w, 75.10.Jm, 82.20.Mj
\parbox[t]{12.5cm}{}
\end{flushleft}
\normalsize
\thispagestyle{empty}
\mbox{}
\pagestyle{plain}
%
%
%
\newpage
\setcounter{page}{1}
\setcounter{equation}{0}
Even for complicated one-dimensional many-particle models,
the ground state can have a simple form. In spin problems
it may be the tensor product of factors referring to
single sites. While the correlations in this case are
trivial, this is not so for a generalization where the product
state is formed using matrices \cite{Type1}-\cite{ThreeState}.
It has been found that the ground state of certain spin-one
models and the stationary state for classical particles
diffusing between two reservoirs have such a form. Also excited states
have been described by the same Ansatz \cite{Stinchcombe}. So far,
however, only diffusive  systems could be treated successfully
in this
way. It is the aim of the present letter to show that the approach
also works for more general situations. As an example, a
particular reaction-diffusion model will be studied.
\\
\indent
We consider a stochastic two state model on a one-dimensional
lattice with $N$ sites. Its configurations are defined
by the occupation numbers $\tau_1,\tau_2,\ldots,\tau_N\;$
each of which can take values $0$ and $1\;$.
We say the system has a matrix product ground
state if its stationary probability distribution
$P_0(\tau_1,\tau_2,\ldots,\tau_N)$ can be written as
\vspace{-2mm}
\begin{eqnarray}
\label{MatrixAnsatz}
P_0(\tau_1,\tau_2,\ldots,\tau_N) \;&=&\;{Z_N}^{-1}\;
\langle W| \,
\prod_{j=1}^N (\tau_j D \,+\, (1-\tau_j)E\,)\,
|V\rangle
\end{eqnarray}
where $E$ and $D$ are square matrices and $\langle W|$
and $|V\rangle$ are vectors acting in an auxiliary space.
$Z_N$ is a normalization constant defined as
$Z_N=\langle W|(D+E)^N |V\rangle$.
The matrix product in Eq. (\ref{MatrixAnsatz})
can be written formally as a tensor product so that
the stationary state $|P_0\rangle$ represented as a vector
in configuration space is given by
\begin{equation}
\label{GroundState}
|P_0\rangle \;=\;{Z_N}^{-1}\; \langle W| \,
\left( \hspace{-1.5mm} \begin{array}{c} E \\[-1mm]
 D \end{array} \hspace{-1.5mm}  \right) ^ {\otimes N}
|V\rangle\,\,.
\end{equation}
The matrix product representation is a powerful tool since
various physical quantities like the particle density
\begin{equation}
\label{ParticleDensity}
\langle \tau_j \rangle_N \;=\;
\frac{\langle W | \, C^{j-1} D C^{N-j} \, | V \rangle}
     {\langle W | \, C^N \, | V \rangle}
\hspace{10mm}
\mbox{with}
\hspace{5mm}
C=D+E\,.
\end{equation}
can be computed directly. Correlation functions
are given by similar expressions in which $C$ plays
the role of a transfer matrix.
\\
\indent
The matrices used for the above Ansatz may by
finite or infinite dimensional \cite{Type2,vladimir}.
We are going to study an example for the first case below.
The fact that the probability distribution of
some system is given by a  product of finite
dimensional matrices has far reaching consequences.
Depending on the properties of the matrix
$C$, correlation functions in such systems
can have two forms which we want to discuss briefly at this point.
Suppose first that the $d$-dimensional matrix
$C$ is diagonalizable and has eigenvalues $\lambda_1,...,\lambda_d$ with
 $\lambda_1 \le \lambda_2 \le...\le \lambda_{d-1} \le \lambda_d\;$.
Then any $n$-point correlation function can be written as
\begin{eqnarray}
\langle \tau_{j_1} \tau_{j_2}...\tau_{j_n}\rangle_N &=&
\sum_{ \{\sigma_i^{\mu}\} }
\;c_N( \{\sigma_i^{\mu} \})\;
\exp\;\{\;-\sum_{\mu=1}^{d-1}\;(\xi_{\mu})^{-1}\;[\;
(j_1-1) {\sigma_1^{\mu}}+(j_2-j_1-1) {\sigma_2^{\mu}}\nonumber\\
\label{correlation1}
&&+(j_3-j_2-1) {\sigma_3^{\mu}} +...+
(j_n-j_{n-1}-1) {\sigma_n^{\mu}} +
(N-j_n) {\sigma_{n+1}^{\mu}}\;]\;\}
\;\;\end{eqnarray}
where the first sum runs over all $\sigma_i^{\mu}=0,1$ under
the restriction $\sum_{\mu=1}^{d-1}\sigma_i^{\mu} \le 1\;$.
The quantities $\xi_{\mu} = \{\log (\lambda_d/\lambda_{\mu})\}^{-1}$
are the correlation lengths.
The $c_N( \{\sigma_i\})$ are some coefficients which
depend on the system size $N\;$ and approach constant
values for $N \gg 1\;$. All correlations depend
exponentially on the distances involved \cite{vladimir}
and the number of length scales equals the number
of different eigenvalues of $C$ minus one.
\\
\indent
The situation changes if $C$ is not diagonalizable.
In this case the matrix
can  be classified according to its Jordan normal form.
As long as the  Jordan block $J_{max}$ of the largest
eigenvalue is one-dimensional, the correlation functions again
decay exponentially (the only difference to
Eq. (\ref{correlation1}) is that algebraic prefactors
to the exponentials may occur). On the other hand,
if the dimension $l$ of $J_{max}$ is larger than
one, the correlations are dominated by  algebraic terms
with positive powers. One can show easily
that the  $n$-point correlation functions are given by
\begin{eqnarray}
\langle \tau_{j_1} \tau_{j_2}...\tau_{j_n}\rangle_N &=&
\;\sum_{ \{\sigma_i=0,...,l-1\} }\;[\;
c_N( \{\sigma_i\} )\; (j_1-1)^{\sigma_1} (j_2-j_1-1)^{\sigma_2}
(j_3-j_2-1)^{\sigma_3} \nonumber \\
\label{correlation2}
&&
\hspace{15mm}
\cdot ...\cdot
(j_n-j_{n-1}-1)^{\sigma_n} (N-j_n)^{\sigma_{n+1}}\;]
\;+\;\mbox{\em exponential terms}
\;.\end{eqnarray}
Here the $c_N( \{\sigma_i\})$ are coefficients
the large-$N$ asymptotics of which are generically
proportional to $N^{-(l-1)}\;$.
The exponential terms are of type (\ref{correlation1})
with algebraic prefactors and are generally negligible for
large distances $j_1,(j_2-j_1),...,(j_n-j_{n-1}),(N-j_n)\gg1\;$.
The correlations (\ref{correlation2}) are completely different
from those in a system with diagonalizable $C$.
They involve the powers  $0,1,2,...,l-1$ of the distances.
Only in special cases where all the $c_N( \{\sigma_i\})$
vanish the correlations are of type (\ref{correlation1}).
It is worth mentioning another special case:
If $l=2$ and the element $D_{d,d-1}$
of the matrix $D$ is zero, then all $c_N(\{\sigma_i\})$ with more than one
$\sigma_i$ equal to $1$ vanish, i.e. the  correlation functions
are linear in the positions $j_1,\ldots,j_n$. If in addition $D_{d,d}=0$,
the  algebraic part of the correlation function depends on the arguments
$(N-j_n)$ and $N$ only, i.e.
$\langle \tau_{j_1} \tau_{j_2}...\tau_{j_n}\rangle_N \approx c'
(1-j_n/N)+$ {\em exponential terms} \hspace{0.2cm}
for $N\gg 1$ and $(N-j_n)\gg1\;$ where $c'$ is some constant.
Let us also note that  the stationary correlations
of any system with a ground state (\ref{MatrixAnsatz})
containing finite dimensional matrices do not involve
negative or non-integer powers of the distances.

%
%
%
%
Up to now, matrix product ground states have been encountered in two
situations. The first is found for models with the Hamiltonian
$H=\sum_{j} h_{j,j+1}$ in which
the two-site interaction $h$ itself already
annihilates the ground state, i.e. $h_{j,j+1} |0>=0$. Here
the algebra of the operators $E$ and $D$ is given by
\begin{equation}
h \, \left[ \,
\left( \hspace{-1.5mm}  \begin{array}{c} E \\[-1mm]
 D \end{array} \hspace{-1.5mm}  \right) \otimes
\left( \hspace{-1.5mm} \begin{array}{c} E \\[-1mm]
 D \end{array} \hspace{-1.5mm}  \right)
\, \right] \;=\; 0 \,.
\end{equation}
An example for this type of models is the
spin-$1$ antiferrromagnet discussed in Ref. \cite{Type1}.
The second case is realized in models with open boundaries
and particle input and output at the ends of the chain.
These models are described by a
time evolution operator $H=\sum_{j=1}^{L-1} h_{j,j+1}
+ h^{(L)}_1 + h^{(R)}_L$ where $h^{(L)}$ and $h^{(R)}$ are
$2 \times 2$ matrices for particle input and output.
Here the basic mechanism of the matrix product ground state
relies on the fact that application of $h_{j,j+1}$ yields a
divergence-like term on the right hand side
\begin{equation}
\label{OldAnsatzBulk}
h \, \left[ \,
\left( \hspace{-1.5mm} \begin{array}{c} E \\[-1mm]
 D \end{array} \hspace{-1.5mm} \right) \otimes
\left( \hspace{-1.5mm} \begin{array}{c} E \\[-1mm]
 D \end{array} \hspace{-1.5mm} \right)
\, \right] \;\;=\;\;
\left( \hspace{-1.5mm} \begin{array}{c} e \\[-1mm]
 d \end{array} \hspace{-1.5mm} \right) \otimes
\left( \hspace{-1.5mm} \begin{array}{c} E \\[-1mm]
 D \end{array} \hspace{-1.5mm} \right) -
\left( \hspace{-1.5mm} \begin{array}{c} E \\[-1mm]
 D \end{array} \hspace{-1.5mm} \right) \otimes
\left( \hspace{-1.5mm} \begin{array}{c} e \\[-1mm]
 d \end{array} \hspace{-1.5mm} \right)
\end{equation}
where $e$ and $d$ are numbers, normally $e=-d=1$. Summing over
the two-particle interactions, all these contributions cancel in the bulk
of the chain. The remaining terms at the boundaries are canceled by
a proper choice of the vectors $\langle W|$ and $|V \rangle$:
\begin{equation}
\label{OldSurfaceTerms}
<W| \, h^{(L)} \left( \hspace{-1.5mm} \begin{array}{c} E \\[-1mm]
 D \end{array} \hspace{-1.5mm} \right) =
-<W| \, \left( \hspace{-1.5mm} \begin{array}{c} e \\[-1mm]
 d \end{array} \hspace{-1.5mm} \right)\,,
\hspace{10mm}
h^{(R)} \left( \hspace{-1.5mm} \begin{array}{c} E \\[-1mm]
 D \end{array} \hspace{-1.5mm} \right) \, |V> =
\left( \hspace{-1.5mm} \begin{array}{c} e \\[-1mm]
 d \end{array} \hspace{-1.5mm} \right)\, |V>
\end{equation}
so that $H|0\rangle=0$. The most important two-state model of this type
is the asymmetric exclusion process with external particle input and output
\cite{Type2}-\cite{vladimir},\cite{Stinchcombe}.
There are also three-state models to which the
matrix product Ansatz has been applied \cite{ThreeState}.
But, as mentioned in the beginning, all known examples are
 diffusive systems.

The generalization which we are going to use, consists in replacing
the numbers $e$ and $d$ by matrices
$\bar{E}$ and $\bar{D}$. The idea goes back
to Ref. \cite{Stinchcombe} where the special case
$\bar{E}+\bar{D}=0$ has
been introduced in order to solve the time-evolution
of the asymmetric diffusion model in one dimension.
The generalized algebra
\begin{equation}
\label{NewAnsatzBulk}
h \, \left[ \,
\left( \hspace{-1.5mm} \begin{array}{c} E \\[-1mm]
 D \end{array} \hspace{-1.5mm} \right) \otimes
\left( \hspace{-1.5mm} \begin{array}{c} E \\[-1mm]
 D \end{array} \hspace{-1.5mm} \right)
\, \right] \;\;=\;\;
\left( \hspace{-1.5mm} \begin{array}{c} \bar{E} \\[-1mm]
 \bar{D} \end{array} \hspace{-1.5mm} \right) \otimes
\left( \hspace{-1.5mm} \begin{array}{c} E \\[-1mm]
 D \end{array} \hspace{-1.5mm} \right) -
\left( \hspace{-1.5mm} \begin{array}{c} E \\[-1mm]
 D \end{array} \hspace{-1.5mm} \right) \otimes
\left( \hspace{-1.5mm} \begin{array}{c} \bar{E} \\[-1mm]
 \bar{D} \end{array} \hspace{-1.5mm} \right)
\;\;,
\end{equation}
\begin{equation}
\label{NewSurfaceTerms}
<W| \, h^{(L)} \left( \hspace{-1.5mm} \begin{array}{c} E \\[-1mm]
 D \end{array} \hspace{-1.5mm} \right) =
-<W| \, \left( \hspace{-1.5mm} \begin{array}{c} \bar{E} \\[-1mm]
 \bar{D} \end{array} \hspace{-1.5mm} \right)\,,
\hspace{10mm}
h^{(R)} \left( \hspace{-1.5mm} \begin{array}{c} E \\[-1mm]
 D \end{array} \hspace{-1.5mm} \right) \, |V> =
\left( \hspace{-1.5mm} \begin{array}{c} \bar{E} \\[-1mm]
 \bar{D} \end{array} \hspace{-1.5mm} \right)\, |V>
\end{equation}
is quadratic on both the left and the right hand side.
In contrast to the usual matrix Ansatz (\ref{OldAnsatzBulk})
the generalized Ansatz (\ref{NewAnsatzBulk})
can be applied to systems which include particle reactions.
\\[2mm]
\indent
As an example we consider the asymmetric coagulation-decoagulation model.
In this model particles diffuse on a linear chain.
When two of them meet, they can merge (coagulate) to a single one.
In the same way a single particle can split up (decoagulate) into
two particles. Assuming no particle input and output,
we therefore have six different processes:
$$
\begin{array}{rccr}
\mbox{diffusion to the left:} && \emptyset+A \rightarrow A+\emptyset &
\mbox{with rate} \; \; a_L \\
\mbox{diffusion to the right:} && A+\emptyset \rightarrow \emptyset+A &
\mbox{with rate} \; \; a_R \\
\mbox{coagulation at the left:} && A+A \rightarrow A+\emptyset &
\mbox{with rate} \; \; c_L \\
\mbox{coagulation at the right:} && A+A \rightarrow \emptyset+A &
\mbox{with rate} \; \; c_R \\
\mbox{decoagulation to the left:} && \emptyset+A \rightarrow A+A &
\mbox{with rate} \; \; d_L \\
\mbox{decoagulation to the right:} && A+\emptyset \rightarrow A+A &
\mbox{with rate} \; \; d_R
\end{array}
$$
In what follows we consider the special
choice $a_L=c_L=q$, $a_R=c_R=q^{-1}$,
$d_L=\Delta q$ and $d_R=\Delta q^{-1}$
where the diffusion and coagulation rates coincide and
all reactions have the same bias in one spatial direction.
Since in this case the model can be mapped on a free fermion model,
it is integrable and various exact results have been obtained
\cite{Doering,XYEquivalence,OurLastPaper}.
The model is controlled by two parameters, namely the
asymmetry parameter $q$ and the effective decoagulation rate $\Delta$.
Its phase diagram shows two phases, a low-density
phase for $\Delta < q^2-1$
and a high-density phase for $\Delta > q^2-1$.
At the phase transition point $\Delta=q^2-1$, the gap in the
relaxational spectrum vanishes
and algebraic long-range  correlations can be observed
\cite{OurLastPaper}. In a basis
$(\emptyset\emptyset,\emptyset A,A\emptyset,AA)$
the two-site term in the time evolution operator
$H=\sum_{j=1}^{N-1} h_{j,j+1}$ reads
\begin{equation}
\label{Hamiltonian}
h \;=\; \left(
\begin{array}{cccc}
0 & 0 & 0 & 0 \\
0 & (\Delta+1) q & -q^{-1} & -q^{-1} \\
0 & -q & (\Delta+1) q^{-1} & -q \\
0 & -\Delta q & -\Delta q^{-1} & q+q^{-1}
\end{array} \right)\,.
\end{equation}
Therefore the bulk algebra (\ref{NewAnsatzBulk}) is given by
\begin{eqnarray}
\label{BulkAlgebra}
0 &=& \bar{E}E-E\bar{E} \\
(\Delta+1) q \,ED \,-\, q^{-1} DE \,-\, q^{-1} DD &=&
\bar{E}D-E\bar{D} \\
-q\,ED \,+\, (\Delta+1) q^{-1} DE \,-\, q\,DD &=&
\bar{D}E-D\bar{E}\\
\label{BulkAlgebra4}
-\Delta q \, ED \,-\, \Delta q^{-1} DE \,+\, (q+q^{-1})\,DD &=&
\bar{D}D-D\bar{D}
\end{eqnarray}
and the boundary conditions (\ref{NewSurfaceTerms}) read
\begin{equation}
\label{BoundaryRelations}
\langle W | \, \bar{E} = \langle W | \, \bar{D} =
\bar{E} | V \rangle = \bar{D} | V \rangle \;=\; 0\,.
\end{equation}
Writing $C=E+D$, \ \ $\bar{C}=\bar{E}+\bar{D}$ and $\gamma^2=\Delta+1$,
the algebra (\ref{BulkAlgebra})-(\ref{BulkAlgebra4}) simplifies to
\begin{eqnarray}
\label{SimplifiedBulkAlgebra}
&&[C,\bar{C}] \;=\; [E,\bar{E}] \;=\; 0 \\
&&\bar{E}C-E\bar{C} \;=\;
(\gamma^2 q + q^{-1}) \, EC \,-\, \gamma^2 q \, EE  \,-\, q^{-1}\, CC \\
&&\bar{C}E-C\bar{E} \;=\;
(\gamma^2 q^{-1} + q) \, CE \,-\, \gamma^2 q^{-1}\,EE \,-\, q\, CC\,.
\label{SimplifiedBulkAlgebraEnd}
\end{eqnarray}
%
%
%
%
%
%
In contrast to algebras for diffusive systems
(\ref{OldAnsatzBulk})-(\ref{OldSurfaceTerms}), the above
commutation relations do not allow to reduce the number of
factors in a given product of matrices.
Therefore products of different lengths are independent.
Products of the same length, which correspond to a given system size,
obey linear relations as follows.
 For
a given product, e.g.~$CECEE$, we compute the difference
$\bar{C}ECEE-CECE\bar{E}$ by using the commutation relations
(\ref{SimplifiedBulkAlgebra})-(\ref{SimplifiedBulkAlgebraEnd}).
Writing $\langle W|\ldots|V\rangle \equiv \langle\ldots\rangle$
and using $\langle\bar{C}ECEE\rangle =\langle CECE\bar{E}\rangle=0$ one
obtains:
\begin{eqnarray}
\label{Example}
&& \langle CECEE \rangle
  \;=\;
\Bigl(\gamma^2 q^{-1} + \gamma^2 (q+q^{-1}) +
(q+q^{-1}) + q\Bigr)^{-1}\,\Bigl[
 \gamma^2 q^{-1}
\langle EECEE \rangle \\ &&
\hspace{27mm}\;+\;
        \gamma^2 (q+q^{-1}) \,	\langle CEEEE \rangle \,+\,
        (q+q^{-1}) \,		\langle CCCEE \rangle \,+\,
        q	\,		\langle CECCE \rangle \nonumber
\,\Bigr]
\,.
\end{eqnarray}
In general, if $\{P_i^{(k)}, \ i \in 1\ldots N_k\}$
is the subset of products with $N$ factors containing $k$ matrices~$C$,
linear relations of this type have the form
\begin{equation}
\label{Application}
\langle P_i^{(k)} \rangle \;=\;
\sum_{j=1}^{N_{k+1}} c_{i,j}^{k,k+1} \langle P_j^{(k+1)} \rangle
\,+\,
\sum_{j=1}^{N_{k-1}} c_{i,j}^{k,k-1} \langle  P_j^{(k-1)} \rangle
\,.
\hspace{15mm}
(k=1\ldots N-1)
\end{equation}
As can be seen from the commutation relations,
the coefficients $c_{i,j}^{k,k \pm 1}$ obey
\begin{equation}
0 \le c_{i,j}^{k,k \pm 1} \le 1 \,, \hspace{15mm}
0 < \sum_{j=1}^{N_{k \pm 1}} c_{i,j}^{k,k \pm 1} < 1 \,, \hspace{15mm}
\sum_{m=\pm 1}\sum_{j=1}^{N_{k+m}} c_{i,j}^{k,k \pm m} = 1\,.
\end{equation}
Therefore by iterating Eq. (\ref{Application}) one gets
more and more complicated linear expressions with positive
coefficients which involve all subsets $k=0\ldots N$. Since there are
no such relations for $k=0$ and $k=N$, one finally ends up with only two
contributions:
\begin{equation}
\label{InfiniteIteration}
\langle P_i^{(k)} \rangle \;=\;
a_i^{(k)} \langle E^N \rangle
\,+\, (1-a_i^{(k)})\, \langle C^N \rangle\,.
\hspace{15mm}
(0 < a_i^{(k)} < 1)
\end{equation}
%
%
The expectation values $\langle E^N \rangle$ and $\langle C^N \rangle$
are independent. Therefore the vector space of words $P_i^{(k)}$ of a
given length
decomposes into two subspaces in  which the expectation values
are proportional to $\langle E^N \rangle$ or $\langle C^N \rangle\;$,
respectively.
 Consequently physical observables are
parameterized by the ratio
$\lambda :=  \langle E^N \rangle / \langle C^N \rangle$.
This is related to the fact that the
model has two independent
ground states, a trivial one which is the empty
lattice ($\lambda=1$) and a nontrivial one
where particles are present ($\lambda=0$).

%
%
%
%

A trivial representation of the above algebra
(\ref{BoundaryRelations})-(\ref{SimplifiedBulkAlgebraEnd}) is
$E=C=1,\;\bar{E}=\bar{C}=0$ which describes a system without particles.
In the symmetric case $q=1$ there also is a second one-dimensional
representation $E=1,\;C=\gamma^2,\;\bar{E}=\bar{C}=0$ corresponding to
a factorized ground state with finite particle density $\Delta/(1+\Delta)$.
In the general case $q \ne 1$ the model is known to involve
three different length scales, and therefore any nontrivial
representation of the algebra has a dimension $d\ge 4$.
Furthermore representations of the algebra may be different in each sector
so that they may depend explicitly on $N$ and $\lambda$.
We found a four-dimensional representation which is given by
\begin{equation}
E_1 \;=\;
\left(\begin{array}{cccc}
q^{-2} & q^{-2} & 0 & 0 \\
0 & \gamma^{-2} & \gamma^{-2} & 0 \\
0 & 0 & 1 & q^2 \\
0 & 0 & 0 & q^2
\end{array}\right)
\,,
\hspace{20mm}
C_1 \;=\;
\left(\begin{array}{cccc}
q^{-2} & q^{-2} & 0 & 0 \\
0 & 1 & 1 & 0 \\
0 & 0 & \gamma^2 & q^2 \\
0 & 0 & 0 & q^2
\end{array}\right)
\end{equation}
$$
\bar{E}_1 =
\left(\begin{array}{cccc}
0 & 0 & q^{-1} & (q^{-1}-q)^{-1} \\
0 & 0 & q-q^{-1} & -q \\
0 & 0 & \Delta(q-q^{-1}) & -\Delta q \\
0 & 0 & 0 & 0
\end{array}\right)\,,
\hspace{6mm}
\bar{C}_1 =
\left(\begin{array}{cccc}
0 & -\Delta q^{-1} & q^{-1} & (q^{-1}-q)^{-1} \\
0 & \Delta (q^{-1}-q) & q-q^{-1} & -q \\
0 & 0 & 0 & 0 \\
0 & 0 & 0 & 0
\end{array}\right)
$$
\vspace{-1mm}
$$
\langle W_1 | \;=\; \Bigl(
1-q^2,\; 1,\; 0,\; a \Bigr)
\,,
\hspace{15mm}
| V_1 \rangle \;=\; \Bigl(
b,\; 0,\; q^2,\; q^2-1 \Bigr)\,,
$$
\vspace{-2mm}
where
\begin{eqnarray}
\label{ab}
q^{2N}a-q^{-2N}b
&=&\frac{q^{2N}(q^2-\gamma^2)+(\gamma)^{-2N} (\gamma^2-1)(q^2+1) -
        q^{-2N} (q^2 \gamma^2-1)}
       {(\gamma^2-q^2)(\gamma^2-q^{-2})(q^2-q^{-2})} \\&&\;\;+\;
       \frac{\lambda}{\lambda-1}\,\,\frac{(\gamma^2-1)(q^2+1)
        (q^{2N}+q^{-2N}-\gamma^{2N}-\gamma^{-2N})}
       {(\gamma^2-q^2)(\gamma^2-q^{-2})(q^2-q^{-2})}
\,. \nonumber
\end{eqnarray}
%
%
%
%
{\em The case $\Delta \neq q^2-1\;$:}
For practical purposes it is desirable to have a representation in which
the matrix $C$ is diagonal. For $\Delta \neq q^2-1$
an appropriate similarity transformation yields
\begin{eqnarray}
E_2 &=&
\left(\begin{array}{cccc}
q^{-2} & q^2-\gamma^{-2} & q^2-1 & q^2(1-\gamma^2) \\
0 & \gamma^{-2} & 0 & \gamma^2-q^2 \\
0 & 0 & 1 & \gamma^2(q^2-1) \\
0 & 0 & 0 & q^2
\end{array}\right)\,,
\hspace{10mm}
C_2 \;=\;
\left(\begin{array}{cccc}
q^{-2} &&& \\
& 1 && \\
&& \gamma^2 &  \\
&&& q^2
\end{array}\right) \nonumber \\[1mm]
\langle W_2 | &=&
\Bigl(
\frac{1}{1-q^2 \gamma^2}\,,\;\;
0\,,\;\;
\frac{q^2}{q^2 \gamma^2-1}\,,\;\;
\frac{a\,(q^2-q^{-2})(\gamma^2-q^2)\gamma^2 \,-\, q^2\gamma^2}
      {(\gamma^2-1)(q^2+1)}
\Bigr) \\[1mm] \nonumber
| V_2 \rangle &=&
\Bigl(
\frac{b\,(q^4-1)(q^2\gamma^2-1)\,+\,q^4}{q^2+1}
 \,,\;\;
0\,,\;\;
\frac{q^2(\gamma^2-1)}{\gamma^2-q^2}\,,\;\;
\frac{(\gamma^2-1)q^2}{\gamma^4-\gamma^2q^2}
\Bigr)\,.
\end{eqnarray}
%
%
%
%
Using this representation, it is easy to derive
the particle density (\ref{ParticleDensity}) in the sector $\lambda=0$
\begin{equation}
\langle  \tau_j \rangle_N \;=\; \frac{
\gamma^{2N} \Bigl( (\gamma^2-1)+(q^2-1)\gamma^2(q\gamma)^{-2j} \Bigr) -
q^{2N}\Bigl((\gamma^2-1)q^{2-4j}+(q^2-1)(q/\gamma)^{-2j}\Bigr)}
{\gamma^2\,(\gamma^{2N} + \gamma^{-2N}-q^{2N}-q^{-2N}) }
\end{equation}
which coincides with the result obtained in Ref. \cite{OurLastPaper}.
We also checked that the two-point correlation function
$\langle \tau_i \tau_j \rangle_N$ is obtained correctly.
%
%
%
%
\\[1mm]
{\em The case $\Delta=q^2-1\;$ ($q>1)\;$:}
Here the two largest eigenvalues of the
 matrix $C$, namely $1+\Delta$ and $q^2$,
coincide and $C$ is not diagonalizable.
We therefore choose a representation where $C$ has
Jordan normal form
\begin{equation}\label{b3}
E_3=
\left( \begin{array}{cccc}
q^{-2}\,&\, q^{2}-q^{-2}\,&\, q^2-1\,&\,  \frac{q^4+q^{-2}-2q^2}{q^4-1}\\
0\,&\,  q^{-2}\,&\,   0\,&\,   q^{-2}\\
0\,&\,  0\,&\,1\,&\,  -\frac{1+2q^2}{1+q^2}\\
0\,&\,  0\,&\,  0\,&\,  q^2\;
\end{array}\right)\,,
\hspace{5mm}
C_3=
\left( \begin{array}{cccc}
q^{-2}\,&\,0\,&\,0\,&\,0\\
0\,&\,1\,&\,0\,&\,0\\
0\,&\,0\,\,&\,q^2\,&\,1\\
0\,&\,0\,\,&\,0\,&\,q^2\;
\end{array}\right)
\end{equation}
$$
\langle W_3 | = \Bigl( q^{-2}-1\,,\; 0 \,,\; q^2-1 \,,\;
1+a(1-q^{-2}) \Bigr)\,,
\hspace{10mm}
| V_3 \rangle = \Bigl( 1+b(q^2-q^{-2})\,,\; 0\,,\; -1 \,,\; q^2-q^{-2} \Bigr)
$$
%
%
where
%
%
\begin{equation}
q^{2N}a-q^{-2N}b \;=\; \frac{L\,\lambda (q^{2L}-q^{-2L})}{1-\lambda}
\,-\, L\,q^{-2L} \,-\, \frac{q^{2L}-q^{-2L}}{q^{2}-q^{-2}}\,.
\end{equation}
Using this representation, the density
at a site $j$ in the sector $\lambda=0$ is easily obtained as
\begin{equation}\label{b9}
\langle \tau_j \rangle_N \;=\;
\frac{q^{4N}}{q^{4N}-1}\;\left\{
\frac{1}{N}\;+\;\left(1-q^{-2}\right)\left(1-\frac{j}{N}\right)\;+\;
q^{-4 j}
\left[\;\left(q^2-1\right)\left(1-\frac{j}{N}\right)-\frac{1}{N} \right]
\right\}
\;\;
\end{equation}
which agrees with the result from Ref. \cite{OurLastPaper}.
There is no term proportional to $(j-1)$ or $(j-1)(N-j)$ because
of $D_{4,3}=D_{4,4}=0$ (see Eq. (\ref{b3})).
It turns out that any $n$-point correlation function
$\langle \tau_{j_1} \tau_{j_2}...\tau_{j_n}\rangle_N$ depends only on
the two positions  $j_1\,$ and $j_n\;$. According to our discussion at the
beginning
of this letter, its algebraic part is a linear function in $j_n$
only. In fact it is given by
$\frac{q^{4N}}{q^{4N}-1} \left(1-q^{-2}\right)^{n-1}\;\left\{
\left(1-q^{-2}\right)\left(1-\frac{j_n}{N}\right)\;+\;\frac{1}{N} \right\}\;$.
The   exponential part of
$\langle \tau_{j_1} \tau_{j_2}...\tau_{j_n}\rangle_N$
decays with $j_1$ on length scales  $(2 \log q)^{-1}\;$, $(4 \log q)^{-1}$
and with $j_n$ on the length scale  $(2 \log q)^{-1}\;$.

%
%
While an Ansatz of type (\ref{MatrixAnsatz}) with an algebra
(\ref{NewAnsatzBulk}), (\ref{NewSurfaceTerms}) can be made
for any one-dimensional reaction-diffusion model, it is not clear under
which conditions a matrix representation really exists. In particular,
we do not know if the existence of representations is related to the
integrability of the system. One should therefore
investigate non-integrable examples.
Also the extension to systems with open boundaries
would be of interest. However, since some open systems
are known to have correlations decaying with negative powers
of the positions, the corresponding matrix representations are
expected to be infinite-dimensional.

%
%
%
%
\noindent
{\bf Acknowledgments}\\[1mm]
H. H. gratefully acknowledges financial support by the Minerva foundation.
We would like to thank V. Rittenberg and
G. M. Sch\"utz for helpful hints and interesting discussions.
%
%
%
%
%


\vspace{-3mm}
\small
\begin{thebibliography}{99}
\bibitem{Type1}
	V. Hakim (1987) {\it J. Phys.} {\bf A 16} (1983) L213; \\
	A. Kl\"umper, A. Schadschneider and J. Zittartz,
                {\it J. Phys.} {\bf A 24} (1991) L955\\
                and {\it Europhys. Lett.} {\bf 24} (1993) 293
\bibitem{Type2}
	B. Derrida, M. R. Evans, V. Hakim and V. Pasquier,
		{\it J. Phys.} {\bf A 26} (1993) 1493
\bibitem{sa}
	S. Sandow, {\it Phys. Rev. } {\bf E 50} (1994) 2660
\bibitem{vladimir}
	F. H. L. Essler and V. Rittenberg, {\em Representations of the
        quadratic algebra and partially asymmetric diffusion
         with open boundaries}, preprint BONN-TH-95-13, cond-mat/9506131
\bibitem{ThreeState}
        B. Derrida, S. A. Janowsky, J. L. Lebowitz and E. R. Speer,
                {\it Europhys. Lett.} {\bf 22} (1993) 651\\
        M. R. Evans,  D. P. Foster, C. Godr\`eche  and D. Mukamel,
                {\it Phys. Rev. Lett.} {\bf 74}  (1995) 208
\bibitem{Stinchcombe}
        R. B.  Stinchcombe and G. M. Sch\"utz,
              {\it Europhys. Lett.} {\bf 29} (1995) 663;\\
 	R. B.  Stinchcombe and G. M. Sch\"utz,
              {\it Phys. Rev. Lett.} {\bf 75} (1995) 140
\bibitem{Doering}
	C. R. Doering and D. ben-Avraham, {\it Phys. Rev.} {\bf A 38}
	 	(1988) 3055\\
	D. ben-Avraham, M. A. Burschka and C. R. Doering,
		{\it J. Stat. Phys.} {\bf 60}, (1990) 695
\bibitem{XYEquivalence}
	F. C. Alcaraz, M. Droz, M. Henkel and V. Rittenberg,
		{\it Ann. Phys.} {\bf 230} (1994) 250\\
	I. Peschel, V. Rittenberg and U. Schultze,
		{\it Nucl. Phys.} {\bf B 430} (1994) 633\\
	K. Krebs, M. Pfannm\"uller, B. Wehefritz and H. Hinrichsen,
     		{\it J. Stat. Phys.} {\bf 78} (1995) 1429
\bibitem{OurLastPaper}
	H. Hinrichsen, K. Krebs and I. Peschel,
          {\em Solution of a one-dimensional diffusion-reaction
           model with spatial asymmetry}, cond-mat/9507141,
	 	to appear in {\it Z. Physik} {\bf B}
\end{thebibliography}
\end{document}